\documentclass[a4paper,11pt]{article}
\usepackage{pos}
\usepackage{amsmath}
\usepackage{graphicx}
\usepackage{subcaption}
\usepackage{hyperref}
\usepackage{url}
\usepackage{tikz}
\usetikzlibrary{decorations.markings,arrows.meta}
\usepackage{float}
\usepackage{comment}

\title{Study on the systematic effects on $b\to c$ inclusive semileptonic decays}

\author[1]{A.~Barone}
\author[2,3]{A.~Elgaziari\footnote{Speaker}}
\author[4,5]{S.~Hashimoto}
\author[4]{Z.~Hu}
\author[2,3,6]{A.~J\"uttner}
\author[4,5]{T.~Kaneko}
\author[4]{R.~Kellermann}

\affiliation[1]{\small PRISMA+ Cluster of Excellence \& Institut f\"ur Kernphysik, Johannes-Gutenberg-Universit\"at Mainz, D-55099 Mainz, Germany}

\affiliation[2]{\small School of Physics and Astronomy,
University of Southampton, Southampton SO17 1BJ, UK}

\affiliation[3]{\small STAG Research Center,
University of Southampton, Southampton SO17 1BJ, UK}

\affiliation[4]{\small High Energy Accelerator Research Organization (KEK),
Ibaraki 305-0801, Japan}

\affiliation[5]{\small School of High Energy Accelerator Science,
SOKENDAI (The Graduate University for Advanced Studies),
Ibaraki 305-0801, Japan}

\affiliation[6]{\small CERN, Theoretical Physics Department,
Geneva, Switzerland}

\emailAdd{ae3e22@soton.ac.uk}

\abstract{We discuss the calculation of the inclusive semileptonic decay for the process $B_s \to X_c \, l\nu_l$ using lattice QCD. This calculation could be decisive in understanding the long-standing tension between inclusive and exclusive determinations of the CKM matrix element, $|V_{cb}|$. In this talk, we investigate the main sources of systematic uncertainty in these decays, including the impact of Jacobi smearing at the source and sink, variations in source–sink separation, and the intrinsic uncertainties of the inclusive reconstruction method itself. In addition, we explain how we can restrict the reconstruction of the inclusive decay rate to just the excited-state contributions. This is achieved by treating the ground-state contributions as an exclusive decay with well-controlled conventional techniques. Systematic effects from the reconstruction then only affect excited-state contributions. Where these are sub-dominant, a suppression of systematic effects is expected. We show results based on Chebyshev reconstruction, which are part of a larger effort towards a first phenomenologically relevant computation of the inclusive decay rate in the continuum and infinite-volume limits.}

\begin{document}
\maketitle

\section{Introduction}

A long standing tension exists between inclusive and exclusive determinations of the CKM matrix element $|V_{cb}|$. As of today, this tension sits at approximately $3\sigma$: 
\begin{align}
    \text{Exclusive:} &\; |V_{cb}| = (39.46 \pm 0.53) \times10^{-3} \\
    \text{Inclusive:} & \;|V_{cb}| = (42.16 \pm 0.51) \times10^{-3}
\end{align}
The exclusive value comes from an average over several lattice-QCD results as calculated in the FLAG report \cite{FlavourLatticeAveragingGroupFLAG:2021npn}. In contrast, the inclusive determination is based on perturbative QCD \cite{Bordone_2021, finauri2024q2momentsinclusivesemileptonic} (as well as different experimental techniques). As different theoretical frameworks are used for the exclusive and inclusive determinations, it is unclear whether this tension is a sign of new physics or perhaps a symptom of the systematic effects introduced by the  differing methods.

However, recent work has demonstrated that inclusive decays can be calculated using lattice QCD based on Chebyshev reconstruction \cite{ Bailas:2020qmv, Hashimoto:2017wqo, Gambino:2020crt, Gambino:2022dvu, Barone:2023tbl} and Backus-Gilbert based approaches \cite{Hansen:2019idp, DeSantis:2025yfm, DeSantis:2025qbb}. By computing the inclusive decay using lattice-QCD, one may be able to shed some light on this $|V_{cb}|$ tension. Accordingly, our aim is to determine the first phenomenologically relevant computation of the inclusive decay for the $B_s \to X_c$ decay using lattice-QCD (where $X_c$ denotes all allowed hadronic states with  strange and charm valence quarks). However, before such results can be reported a rigorous study of their systematic errors must be conducted. Here, we extend recent work done on the systematic effects of the $D_s$ decay to that of the $B_s$ \cite{Kellermann:2024jqg, Kellermann:2025pzt}. We focus on systematic errors associated with contaminations from $B_s$ excited states (Section 4) as well as those intrinsically associated with the Chebyshev reconstruction (Section 5). However, first, we will provide a brief overview of the inclusive decay calculation and introduce the relevant quantities. 

\section{Inclusive semileptonic decays on the Lattice}
    This section aims to provide a very brief explanation of the key quantities in the inclusive semileptonic $B_s \to X_c$ decay. For a more detailed discussion please see other works \cite{Hashimoto:2017wqo,Gambino:2020crt, Gambino:2022dvu, Barone:2023tbl}. We will start from the definition of the decay rate for the inclusive decay.  
    \begin{align}
         \Gamma &\sim \int_{0}^{\boldsymbol{q}^2_{\text{max}}} d\boldsymbol{q}^2 \sqrt{\boldsymbol{q}^2} \bar{X}(\boldsymbol{q}^2), \;\;\;
         \bar{X}(\boldsymbol{q}^2) \equiv \int_{\omega_{\text{min}}}^{\omega_{\text{max}}} d\omega \, k^{\mu\nu}(\boldsymbol{q},\omega)W_{\mu\nu}(\boldsymbol{q},\omega),
         \label{eq:decayrate_intro}
    \end{align}
    where $q$ is the momentum transfer between the initial state $B_s$ and final state $X_c$, $\omega$ is the final-state hadron energy, $k^{\mu\nu}$ derives from the leptonic tensor and is analytically known, while $W^{\mu\nu}$ is the hadronic tensor. 
    \begin{equation}
        W_{\mu\nu}(q) = \frac{1}{2E_{B_s}} \int d^4x\, e^{iqx} \langle B_s| J^{\mu \dagger}(x) J^{\nu} (0)| B_s\rangle,   
    \end{equation}
    where $J_{\mu}$ denotes the flavour changing current: $J_{\mu} = V_{\mu}-A_{\mu}$ where $V_{\mu}=\bar{b} \gamma_{\mu} c$ and $A_{\mu}=\bar{b} \gamma_{\mu} \gamma_5 c$. It can be shown that the hadronic tensor is related to the Euclidean four-point correlation function via a Laplace transform. 
    \begin{equation}
        C_{\mu\nu}(\boldsymbol{q},t) = \int_0^{\infty} d\omega \,e^{-\omega t} W_{\mu\nu}(\boldsymbol{q}, \omega),
    \end{equation}
    where $C_{\mu\nu}(\boldsymbol{q},t)$ can be calculated from a ratio of two- and four-point functions,
    \begin{equation}
        C_{\mu\nu}(\boldsymbol{q},t) = \frac{1}{2M_{B_s}} \int d^3\boldsymbol{x}\, e^{i\boldsymbol{q}\cdot \boldsymbol{x}} \langle B_s|J_{\mu}^{\dagger}(\boldsymbol{x},0) e^{-Ht}J_{\nu}(0) |B_s \rangle.
    \end{equation}
    By rewriting Eq.~\ref{eq:decayrate_intro}, in terms of the Laplace transform of $W_{\mu\nu}$, the expression for the decay rate can be written in terms of the Euclidean four-point function. To do this substitution, one introduces a Heaviside function, $\theta(x)$, to impose kinematic constraints. 
    \begin{equation}\label{eq:decayrate-with-heaviside}
        \bar{X}(\boldsymbol{q}^2, \sigma) \sim \int_{\omega_{\text{min}}}^{\infty} d\omega \, W_{\mu\nu}(\boldsymbol{q},\omega)k^{\mu\nu}(\boldsymbol{q},\omega)\theta_{\sigma}(\omega_{\text{max}}-\omega),
    \end{equation}   
    where the Heaviside is smeared and replaced with a sigmoid function, 
    \begin{equation}\label{sigmoid}
        \theta_{\sigma}(x) = \frac{1}{1+e^{-x/\sigma}},\;\;\; \text{with}\;\; \theta(x) = \lim_{\sigma \to 0}\theta_{\sigma}(x), 
    \end{equation}
    and in the limit of small $\sigma$, we recover the Heaviside. This smearing parameter, $\sigma$, provides a controlled way to investigate the systematics associated with this approximation - see Section 5. The function $k^{\mu\nu}(\boldsymbol{q},\omega)\theta_{\sigma}(\omega_{\text{max}}-\omega)$, which we will refer to as the kernel function, is approximated as a power series in $e^{-\omega}$ and we thus relate $\bar{X}(\boldsymbol{q}^2)$ to $C_{\mu\nu}(\boldsymbol{q},t)$, \begin{equation}\label{xbar_as_sum_of_cmunu}
        \bar{X}(\boldsymbol{q}^2, \sigma, N) \sim \sum_k^N a_{\mu\nu, k}(\boldsymbol{q}, \sigma)\int_{\omega_{\text{min}}}^{\infty} d\omega W_{\mu\nu}(\boldsymbol{q},\omega)e^{-\omega t} = \sum_k^N a_{\mu\nu, k}(\boldsymbol{q}, \sigma) C_{\mu\nu}(\boldsymbol{q},t=k),
    \end{equation} 
    where $a_{\mu\nu, k}$ are the coefficients of this approximation and can be calculated numerically from the projection of the kernel function onto the exponential basis and $N$ is the order of this approximation. It's important to note that the order of this approximation is controlled by the time separation between the two currents of the four-point function. Since the four-point function is calculated from lattice data, it suffers from signal deterioration for large time separations of the currents. In order to regularise the signal, we expand in terms of the Chebyshev basis. These polynomials have the property that they are bounded in the interval $[-1,1]$ which can be included in the inclusive calculation as a Bayesian prior. Using this basis, Eq.~\ref{xbar_as_sum_of_cmunu} is rewritten in terms of objects referred to as the Chebyshev matrix elements.
    \begin{equation}\label{xbar_as_sum_of_chebs}
        \bar{X}(\boldsymbol{q}^2, \sigma, N) \sim \sum_k^N c_{\mu\nu, k}(\boldsymbol{q}, \sigma) \langle T_k\rangle_{\mu\nu}.
    \end{equation} 
    These Chebyshev matrix elements can be build from a linear combination of ${C}_{\mu\nu}(t+2t_0)/{C}_{\mu\nu}(t_0)$, i.e. a normalised form of $C_{\mu\nu}$ with $t_0$ as some reference time. Finally, to recover the relevant form of the differential decay rate, $\bar{X} (\boldsymbol{q}^2)$, one must take the $\sigma \to \infty$ and $N \to \infty$ limits.\footnote{In practise, the infinite volume limit must be performed before the $\sigma\to0$ limit. However, it can be shown that for the $B_s$ decay, by using lattice data on multiple volumes, the finite volume effects are negligible.}

\section{Simulation details}
All the results of the following sections are computed on an ensemble of RBC/UKQCD $N_f=2+1$ domain wall fermion (DWF) field configurations \cite{allton2008} of size $24^3\times64$ with the Iwasaki gauge action \cite{iwasaki1983renormalization}. Simulations were computed with lattice spacing, $a \approx0.11$fm and light quark masses such that $M_{\pi} \approx330$MeV. For the valence charm and bottom quarks, we use M\"obius DWF \cite{ Brower:2012vk} and the relativistic heavy quark (RHQ) action \cite{Christ:2006us}, respectively. All the correlation function analysed in this work were generated with \texttt{Grid} \cite{BoyleGrid} and \texttt{Hadrons} \cite{PortelliHadrons} using DiRAC HPC resources.

\section{Systematic errors of the Correlation functions}

\begin{figure}[H]
    \centering
    \begin{tikzpicture}[
        scale=1,
        thick,
        >=Stealth,
        every node/.style={font=\small}
    ]
    
        \coordinate (L)  at (0,0);
        \coordinate (R)  at (6,0);
        \coordinate (U1) at (1.75,0.96);
        \coordinate (U2) at (4.25,0.96);
        
        \coordinate (B1QUARK) at (1,0.65);
        \coordinate (CQUARK) at (3,1.1);
        \coordinate (B2QUARK) at (5,0.65);
        \coordinate (SQUARK) at (3, -1.1);
        
        \tikzset{
          three arrows/.style={
            decoration={
              markings,
              mark=at position 0.20 with {\arrow{>}},
              mark=at position 0.51 with {\arrow{>}},
              mark=at position 0.84 with {\arrow{>}}
            },
            postaction=decorate
          },
          one arrow/.style={
            decoration={
              markings,
              mark=at position 0.51 with {\arrow{<}}
            },
            postaction=decorate
          }
        }
        
        \draw[three arrows] (L) to[out=40, in=140] (R);
        
        \draw[one arrow] (L) to[out=-40, in=-140] (R);
        
        \fill (L)  circle (2pt);
        \fill (R)  circle (2pt);
        \fill (U1) circle (2pt);
        \fill (U2) circle (2pt);
        
        \node[left]  at (L) {$O^{S\dagger}_{B_s}(\boldsymbol{x_{\text{src}}},t_\text{src})$};
        \node[right] at (R) {$O^S_{B_s}(\boldsymbol{x_{\text{snk}}}, t_\text{snk})$};
        
        \node[above] at (U1) {$J_{\nu}(\boldsymbol{x}_1, t_{\text{ins}})$};
        \node[above] at (U2) {$J_{\mu}^{\dagger}(\boldsymbol{x_2}, t_2)$};
    
        \node[below] at (B1QUARK)
        {$b$};
        \node[below] at (CQUARK)
        {$c$};
        \node[below] at (B2QUARK)
        {$b$};
        \node[above] at (SQUARK)
        {$\bar{s}$};  
    \end{tikzpicture}  
    \caption{Schematic quark-flow diagram for the four-point correlation function. The three parameters to be varied are the smearing, denoted as superscript $S$ on the $B_s$ creation and annihilation operators; the distance between $t_\text{src}$ and $t_\text{snk}$; and the location of $t_{\text{ins}}$ which for a given sequential propagator is fixed.}
    \label{fig:propagator}
    \end{figure}
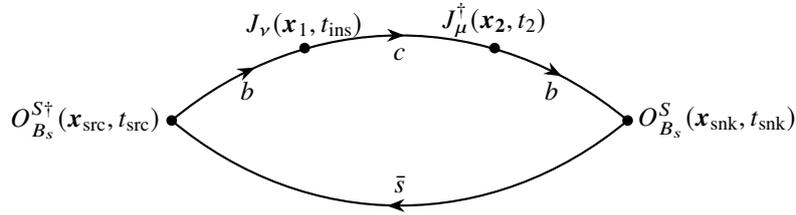

    In this section, we investigate the systematic errors associated with contaminations from $B_s$ excited states. To explore this source of systematic error, the parameters of the four-point function were varied and $\bar{X}$ was calculated in each case. To visualise the parameters of the four-point function, a schematic quark-flow diagram is provided in Fig.~\ref{fig:propagator}. 

\begin{figure}
        \centering
        \includegraphics[width=\linewidth]{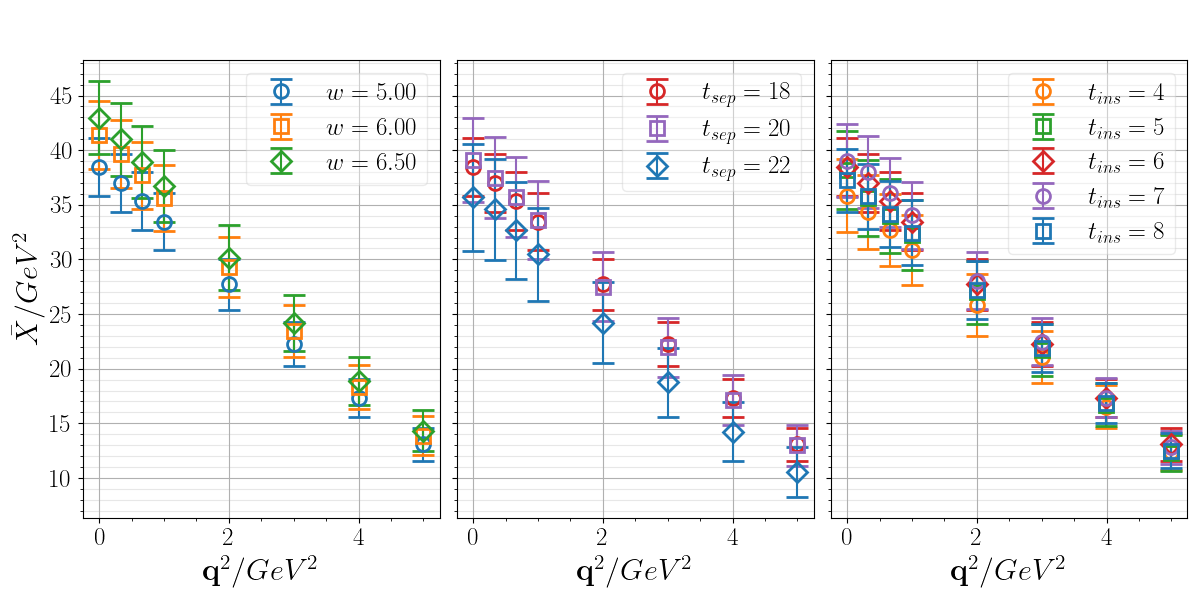}
        \caption{$\bar{X}(\boldsymbol{q}^2)$ with $\boldsymbol{q}^2$ for the full kinematical regime. The different colours refer to the choices of smearing width denoted as $w$ (left); the time separation  of the source and sink in the four-point function in lattice units (middle); and the location of the fixed current time insertion in lattice units (right). All variations of $\bar{X}$ for a given $\boldsymbol{q}^2$ are consistent within error.}
        \label{fig:xbar_all}
    \end{figure}

    \noindent We vary three different parameters to explore this excited-state contamination. This was done for values of $\boldsymbol{q}^2$ across the full kinematic regime. 
    \begin{itemize}\setlength{\itemsep}{0pt}
        \item The \textbf{Jacobi smearing} applied to the source and sink operators \cite{ allton1993gauge}. In the left figure of Fig.~\ref{fig:xbar_all}, we show $\bar{X}(\boldsymbol{q}^2)$ for three values of smearing width, $w$ ($5.0$, $6.0$ and $6.5$). 
        There is a slight trend across these three smearing values. To investigate this further, we computed the four-point function for multiple smearing widths on a lattice with $L=32^3$. On this larger lattice no dependence on the smearing was seen and thus, we conclude that the smearing dependence observed for $L=24^3$ is not significant.         
        \item The \textbf{time separation} between the source and the sink operators, $T_{\text{sep}}=t_\text{snk}-t_\text{src}$. Three choices of $T_{\text{sep}}$ were chosen: $18, 20 \text{ and } 22$ (lattice units). In the middle figure of Fig.~\ref{fig:xbar_all}, we see that, for these three choices, $\bar{X}$ for a given $\boldsymbol{q}^2$ is compatible within errors.
        \item The \textbf{location of the fixed current time} location. To compute multiple values of $t_2-t_{\text{ins}}$, 
        one time insertion of the current is fixed (e.g $t_{\text{ins}}$) and the other is varied (e.g $t_2$). Five different values of $t_{\text{ins}}$ were chosen: $t_{\text{ins}}=4,5,6,7$ and $8$ (lattice units). Displayed in the right figure of Fig.~\ref{fig:xbar_all}, We see that across these values, once again $\bar{X}$ is within error for the different time insertions across the kinematical regime. 
    \end{itemize}

\begin{figure}
        \centering
        \includegraphics[width=1.1 \linewidth,trim=2.5cm 0cm 0cm 0cm,clip]{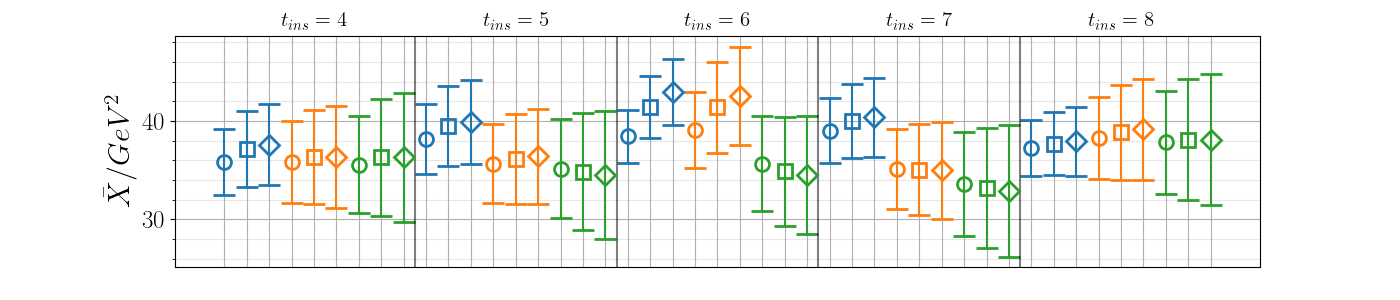}
        \captionsetup{skip=5pt}
        \caption{$\bar{X}(\boldsymbol{q}^2)$ at $\boldsymbol{q}^2=0.73\,$GeV$^2$ for a range of different four-point function parameters. On the top axis, we denote different choices of $t_{\text{ins}}$. The blue, orange and green colours refer to the time separations of source and sink of $18, 20 \text{ and } 22$. Whilst the circle, square and diamond markers denote smearing widths of $5.0, 6.0 \text{ and } 6.5$. All results are roughly within one $\sigma$.}
        \label{fig:xbar-all-systematics}
    \end{figure}

    These parameters of the four-point function were also varied simultaneously. This is shown in Fig.~\ref{fig:xbar-all-systematics}. Here, $\bar{X}$ is shown for one particular value of $\boldsymbol{q}^2$. Across all the different choices of parameters $\bar{X}$ is stable and always consistent within error. This result has also been verified for all the values of $\boldsymbol{q}^2$ in the kinematic regime. The lack of variation across these different parameters is a sign that the contribution of excited $B_s$ states is under control. Going forward, we choose the following values:  $w=6.0$, $T_{\text{sep}}=20$, and $t_{\text{ins}}=6$. These choices sit in the middle of this systematics analysis and have been shown to be stable.  

\section{Smearing dependence of the data}

    In this section, we now discuss the systematic effects associated with the Chebyshev reconstruction - specifically the $\sigma$-smearing dependence of $\bar{X}(\boldsymbol{q}^2)$. Recall that a Chebyshev polynomial approximation was performed on the kernel function, $k_{\mu\nu}(\boldsymbol{q},\omega) \theta_{\sigma}(\omega_{\text{max}}-\omega)$. The smaller the value of $\sigma$ the slower the convergence of the coefficients and thus the more relevant the higher-order terms become. Thus, as $\sigma$ is reduced, $N$ must increase but this increase is limited by the time separation, $t$, of the two currents in $C_{\mu\nu}$. Furthermore, due to the deterioration of the signal of $C_{\mu\nu}$ with increasing $t$, one can suppose that beyond some value of $t$, $t_{\max}$, the signal of $C_{\mu\nu}$ is lost completely to noise. Thus, the values of $\langle T_t\rangle_{\mu\nu}$ for $t>t_{\max}$ determined from $C_{\mu\nu}(t)$ will also just be noise. Beyond the value of $t_{\text{max}}$, we assume the Chebyshev matrix elements to be randomly distributed in the interval $[-1,1]$. In this way, we mimic what we would extract for $t > t_{\text{max}}$ if we were to run lattice simulations for such times. To provide the most conservative estimate for these higher orders, we randomly sample values for $\langle T_{t>t_{\max}}\rangle_{\mu\nu}$ as either $1$ or $-1$ across the bootstrap samples. \begin{equation}\label{eq:xbar_with_highorder_smearing}
        \bar{X} = \underbrace{\sum_t^{t_{\text{max}}} \tilde{c}_{\mu\nu, \, t}(\sigma) \langle T_t\rangle_{\mu\nu}}_{\text{Signal}} +  \underbrace{\sum_{t_{\text{max}}+1}^N \tilde{c}_{\mu\nu, \, t}(\sigma) \cdot \langle T_t \rangle_{\mu\nu}}_{\text{Noise}}.
    \end{equation}
    Of course, the true signal for these higher orders will not be extracted but, by this method, a bound will be generated in which the true signal must exist. Computing $\bar{X}$ in this way allows us to vary $\sigma$ and track the error dependence of $\bar{X}$ on $\sigma$. 

\begin{figure}
        \centering
        \begin{subfigure}{0.48\linewidth}
            \centering
            \includegraphics[width=\linewidth]{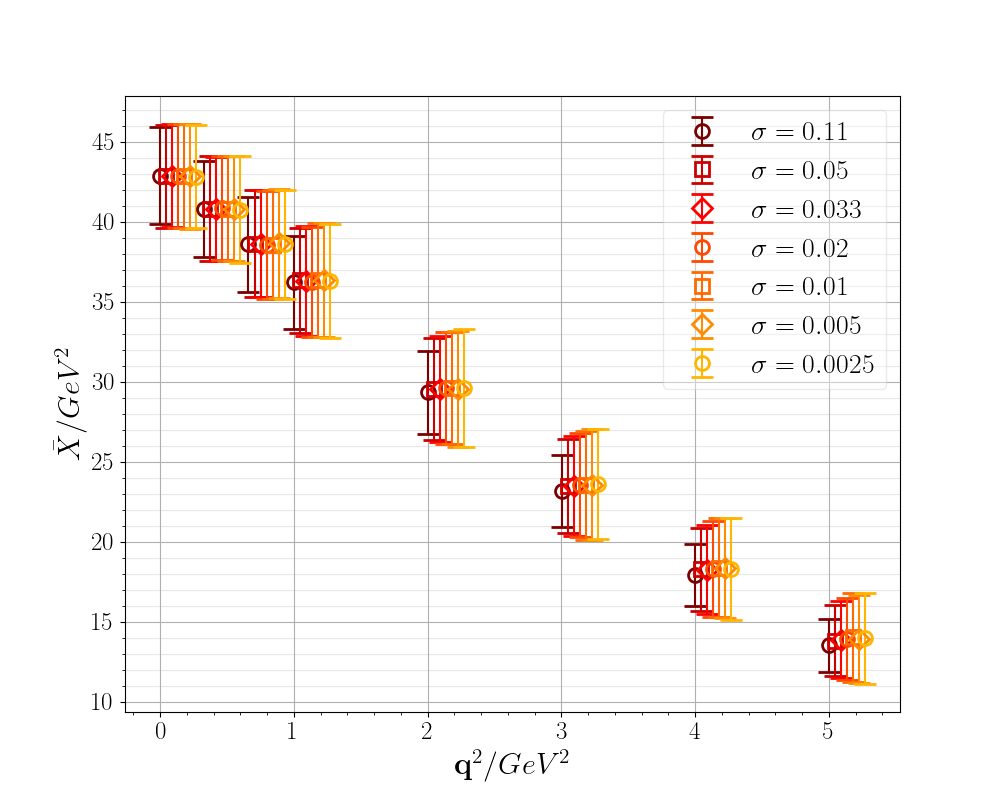}
        \end{subfigure}
        \hspace{0.01\linewidth}
        \begin{subfigure}{0.48\linewidth}
            \centering
            \includegraphics[width=\linewidth]{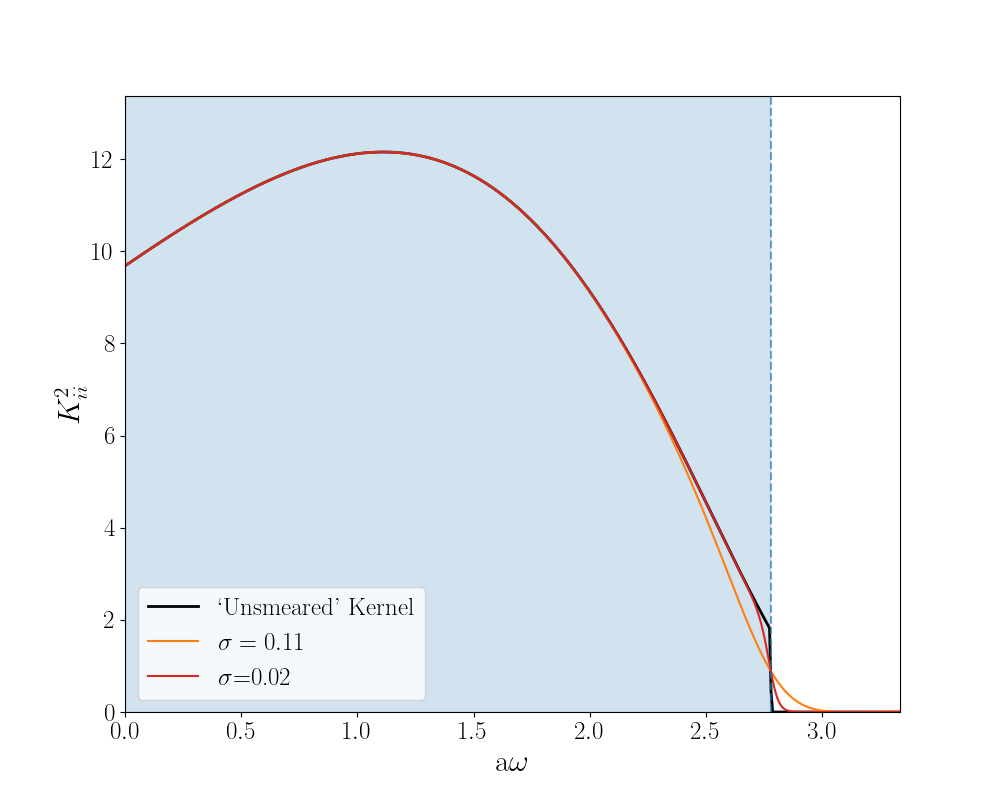}
        \end{subfigure}
        \caption{(Left) Total $\bar{X}$ (summed over all channels) with $\boldsymbol{q}^2$ for several values of the smearing parameter, $\sigma$. Each group of points refer
        to the same value of $\boldsymbol{q}^2$ and have been separated for visual purposes. $\bar{X}$ has been calculated from lattice data using Eq.~\ref{eq:xbar_with_highorder_smearing}. (Right) The form of the kernel function for the dominant channel, $J_{\mu}^{\dagger}=A_i^{\dagger}$ and $J_{\nu}=A_i$, that contributes to $\bar{X}$. The kernel is shown when there is no smearing applied (black) and also two choices of smearing parameter $\sigma = 0.02$ (red) and $\sigma = 0.11$ (orange). Notice how the form of the curves doesn't change much between these choices.}
        \label{fig:xbartot_smearing_dep}
    \end{figure}

    For the total value of $\bar{X}$ (summed over all the different $V-A$ channels) we computed $\bar{X}$ using Eq.~\ref{eq:xbar_with_highorder_smearing}.  The results for $\bar{X}$ at multiple values of $\boldsymbol{q}^2$ for different values of $\sigma$ are shown in Fig.~\ref{fig:xbartot_smearing_dep}.  Whilst at high $\boldsymbol{q}^2$ there is a slight increase in the error of $\bar{X}$, overall, there is a mild dependence  on $\sigma$. The mild dependence on $\sigma$ can be related to the form of the kernel functions that contribute to $\bar{X}$. The dominant channel for this decay occurs when the inserted currents, $J_{\mu}^{\dagger}$ and $J_{\nu}$, are $A_i^{\dagger}$ and $A_i$ (respectively) with spatial index: $i$. For this channel, the kernel function is not particularly sharply peaked as displayed in the right hand figure of Fig.~\ref{fig:xbartot_smearing_dep}. As a result, the coefficients don't change very much as the value of $\sigma$ is reduced and thus we see a mild dependence of $\bar{X}$ with $\sigma$. 

\begin{figure}
            \centering
            \begin{subfigure}{0.48\linewidth}
                \centering
                \includegraphics[width=\linewidth]{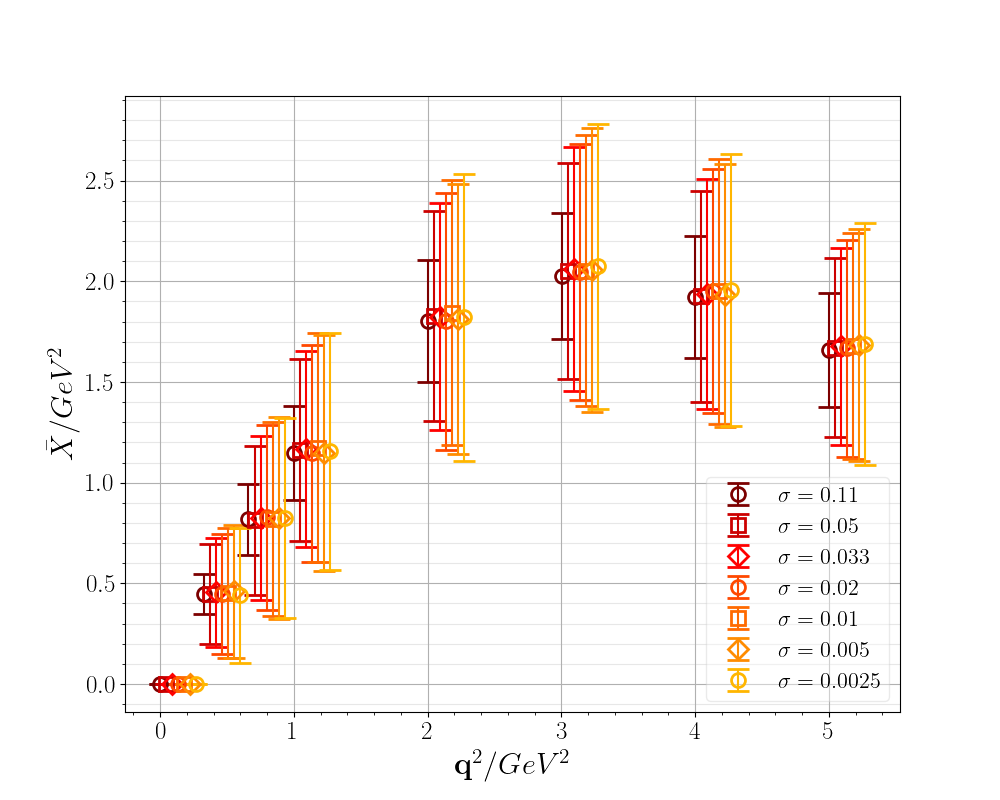}
            \end{subfigure}
            \hfill
            \begin{subfigure}{0.48\linewidth}
                \centering
                \includegraphics[width=\linewidth]{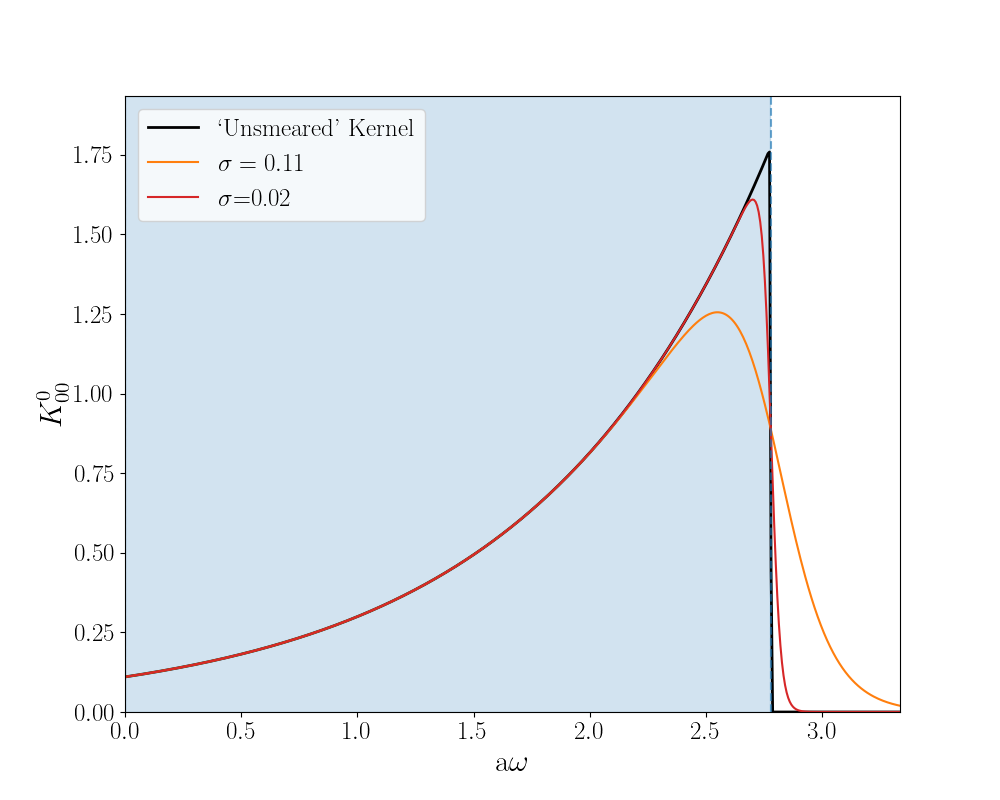}
            \end{subfigure}
            \caption{(Left) Single channel of $\bar{X}$, $J_{\mu}^{\dagger}=V_0^{\dagger}$ and $J_{\nu}=V_0$, with $\boldsymbol{q}^2$ for several values of the smearing parameter, $\sigma$. As in Fig.~\ref{fig:xbartot_smearing_dep}, Each group of points refer to the same value of $\boldsymbol{q}^2$ and $\bar{X}$ has been calculated from lattice data using Eq.~\ref{eq:xbar_with_highorder_smearing}. Now there is a large increase in error of the data as $\sigma$ is reduced. This behaviour can be linked to the form of the kernel function (right). Due to the sharply peaked nature of the kernel function, the smearing approximations for different values of $\sigma$, the orange and red curves, show significant differences.}
            \label{fig:xbarV0V0_smearing_dep}
        \end{figure}

    However, some channels that contribute to $\bar{X}$ have a stronger dependence on $\sigma$. For example, $\bar{X}$ for the subdominant channel, where $J_{\mu}^{\dagger}=V_0^{\dagger}$ and $J_{\nu}=V_0$, shows a strong dependence on $\sigma$ - Fig.~\ref{fig:xbarV0V0_smearing_dep}. And this can be related to its sharply peaked kernel function  - right-hand figure of Fig.~\ref{fig:xbarV0V0_smearing_dep}. Now, as this channel is sub-dominant for the $B_s$ semileptonic decay, this dependence is of limited consequence. On the other hand, if one were to perhaps investigate a different decay, it is possible that the dominant channel could indeed have a strong $\sigma$ dependence. If this is the case, the following section outlines a method to potentially reduce the systematic error associated with this strong $\sigma$ dependence. 
        
    \subsection{Ground-state separation}

        The error increase in the $\sigma \to 0$ limit can be mitigated by separating the ground-state and excited-state contributions to $C_{\mu\nu}$ for each channel \cite{Kellermann:2025pzt}. 
        \begin{equation}
            C_{\mu\nu} = C_{\mu\nu}^{GS} + C_{\mu\nu}^{ES},
        \end{equation}
        where $C_{\mu\nu}^{GS}$ ($C_{\mu\nu}^{ES}$) is the ground (excited) state contribution to $C_{\mu\nu}$. The ground-state contribution to $\bar{X}$ ($\bar{X}^{GS}$) can be calculated from the $C_{\mu\nu}^{GS}$ using well established exclusive techniques. Whilst the excited-state contribution to $\bar{X}$ ($\bar{X}^{ES}$) can be calculated from $C_{\mu\nu}^{ES}$ using inclusive techniques. As a result, the $\sigma$ dependence only enters in the excited-state calculation. As we expect the ground-state to dominate the contributions to $C_{\mu\nu}$ and thus $\bar{X}$, the absolute error in the $\sigma$ limit should be reduced.

        The following discussion is restricted to the channel with $J_{\mu}^{\dagger}=V_0^{\dagger}$ and $ J_{\nu}=V_0$ which shows strong dependence on $\sigma$. In order to separate $C_{\mu\nu}$ into these two parts, we must reconstruct $C_{\mu\nu}^{GS}$. We identify the ground-state contribution for this channel as the $D_s$ state and thus, the ground-state contribution to the four-point function can be written down in the following way.
        \begin{equation}\label{four-point_gs_contribution}
            C^{GS}_{V_{0}V_{0}}(t) = \frac{1}{4 M_{B_s} E_{D_s}} 
                \langle B_s | V^\dagger_{0} | D_s \rangle 
                \langle D_s | V_{0} | B_s \rangle 
                \, e^{-E_{D_s} t}. 
        \end{equation}
        To compute this term, we use a double-exponential fit on the four-point function and identify the first term as the ground-state term as in Eq.~\ref{four-point_gs_contribution}. For more details on fitting of four-point functions please see other works \cite{Kellermann:2025pzt, Hu:2025hpn, hu2026inclusive}. 

\begin{figure}
        \centering
    
        \begin{subfigure}{0.49\linewidth}
            \centering
            \includegraphics[width=\linewidth]{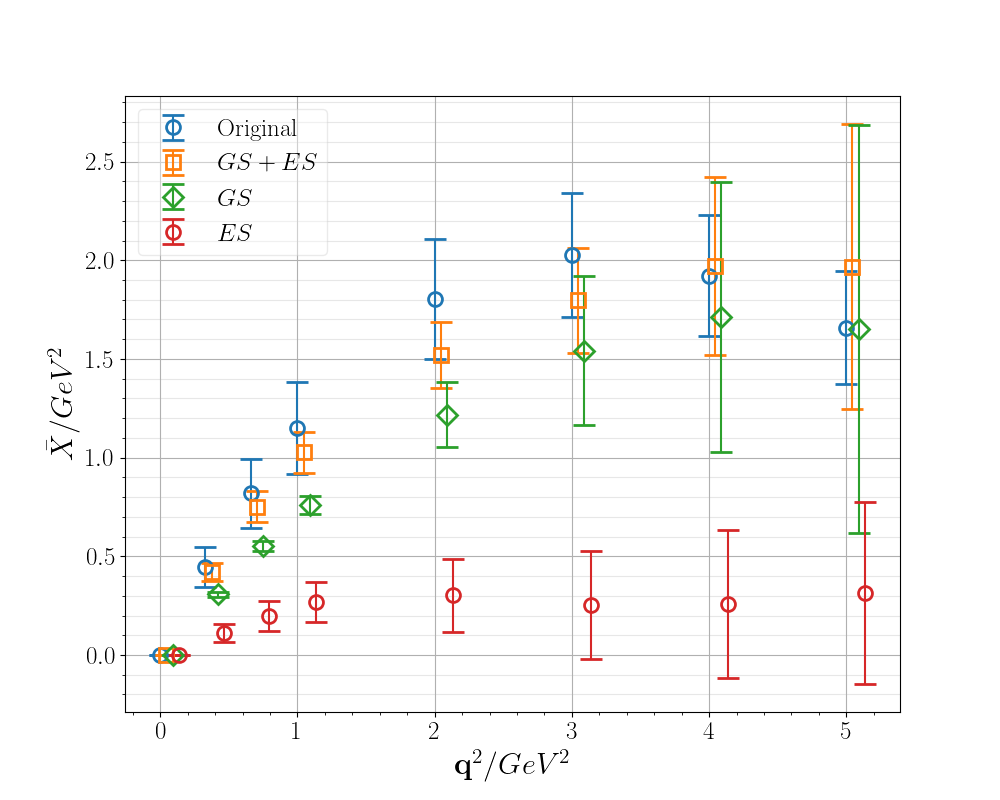}
        \end{subfigure}
        \hfill
        \begin{subfigure}{0.49\linewidth}
            \centering
            \includegraphics[width=\linewidth]{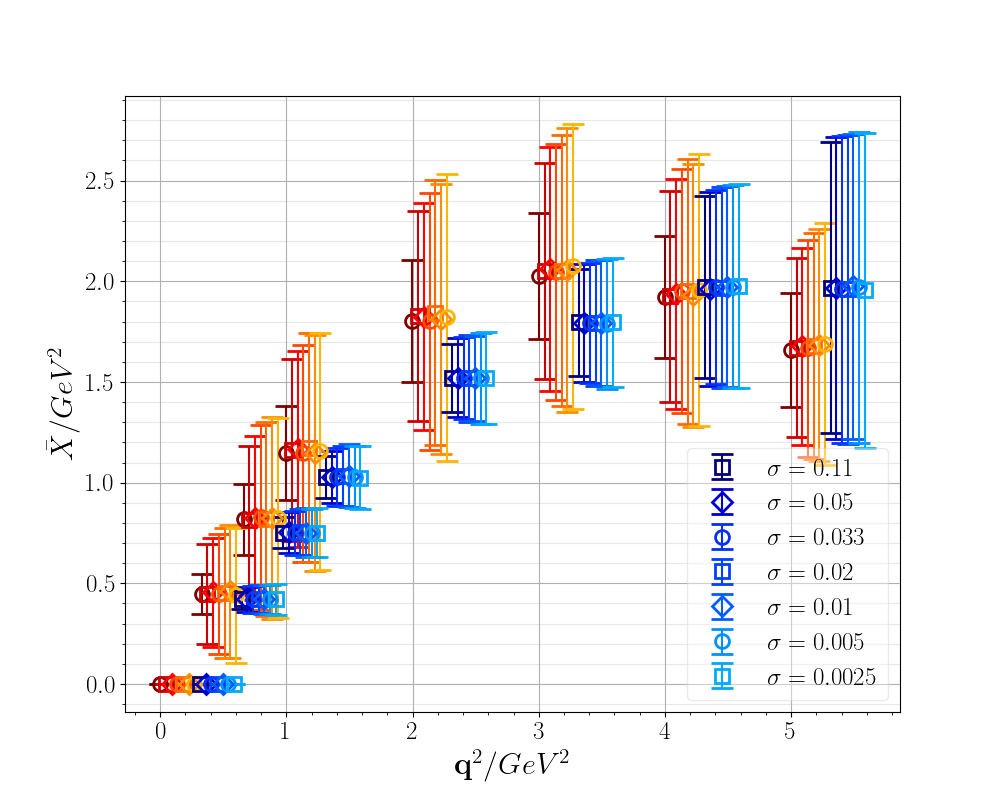}
        \end{subfigure}
    
        \caption{
        $\bar{X}$ for the channel: $J_{\mu}^{\dagger}=V_0^{\dagger}$ and $J_{\nu}=V_0$.
        (Left) Separation into ground- and excited-state contributions:
        blue circles = unseparated, orange squares = $\bar{X}^{GS}+\bar{X}^{ES}$,
        red circles = $\bar{X}^{ES}$, and green diamonds = $\bar{X}^{GS}$.
        (Right) $\bar{X}$ as a function of $\boldsymbol{q}^2$ comparing the original (red)
        and separated-and-summed reconstruction (blue) in the $\sigma\to0$ limit. Notice the reduced dependence on $\sigma$ for the blue points. 
        }
        \label{fig:xbar-combined}
    \end{figure}

        Once the ground-state contribution to the four-point function is computed, $\bar{X}^{GS}$ ans $\bar{X}^{ES}$ can be computed separately. We show these results as well as the values of $\bar{X}$ as obtained from no separation process in the left figure of Fig.~\ref{fig:xbar-combined}. The difference in behaviour with $\sigma$ between the original (red points) and separated-and-summed form of $\bar{X}$ (blue points) can be seen in the right figure of Fig.~\ref{fig:xbar-combined}. As expected, it is clear that the separated-and-summed $\bar{X}$ shows significantly less dependence on $\sigma$ as compared to the original $\bar{X}$ (also shown in Fig.~\ref{fig:xbarV0V0_smearing_dep}). For all values of $\boldsymbol{q}^2$, except the largest, this reduced dependence on $\sigma$ leads to a reduction in the error. For the largest value of $\boldsymbol{q}^2$, the reconstruction of the ground-state is unstable and shows large uncertainties. It may be possible to further constrain these high $\boldsymbol{q}^2$ values using information from the two-point function in the form of a Bayesian prior.  
    
\section{Summary}

    This work has explored the different sources of systematic error in the determination of the inclusive decay rate. We have shown that for some particular choices of smearing width, source sink separation and current insertion that contaminations from $B_s$ excited states can be controlled. In addition, by separating out the ground-state and excited-state contributions to the four-point function, the effects of the smearing limit on the final quantity can be reduced. The methods shown here can be applied to other decays and may provide a systematic way to limit the error in the $\sigma \to 0$  limit. 

\section*{Acknowledgements}

This work used the DiRAC Extreme Scaling service at the University of Edinburgh, operated by the Edinburgh Parallel Computing Centre on behalf of the STFC DiRAC HPC Facility (www.dirac.ac.uk). This equipment was funded by BEIS capital funding via STFC
capital grant ST/R00238X/1 and STFC DiRAC Operations grant ST/R001006/1. DiRAC is part of the National e-Infrastructure. A.E is supported by the Mayflower scholarship in the School of Physics and Astronomy of the University of Southampton. The works of S.H., R.K. and T.K. are supported in part by JSPS KAKENHI Grant Numbers 22H00138, 22K21347, 23K20846 and 25K01007. T.K. is also supported by the U.S.-Japan Science and Technology Cooperation Program in High Energy Physics
(Project ID: 2024-40-2). A.J. is supported by the Eric \& Wendy Schmidt Fund for Strategic Innovation (grant agreement SIF-2023-004).


\setlength{\bibsep}{0pt}

\end{document}